
\input harvmac

\def\p{\partial}
\def\l{\lambda}
\def\scri{{\cal I}}
\input epsf
\noblackbox
\overfullrule=0pt
\def\Title#1#2{\rightline{#1}\ifx\answ\bigans\nopagenumbers\pageno0\vskip1in
\else\pageno1\vskip.8in\fi \centerline{\titlefont #2}\vskip .5in}
 
scaled\magstep3 
 
scaled\magstep3 
 
scaled\magstep3 
 
\font\cmss=cmss10 \font\cmsss=cmss10 at 7pt
\def\ove#1#2{\vbox{\halign{$##$\cr
#1{\leftarrow}\mkern-6mu\cleaders\hbox{$#1\mkern-2mu{-}\mkern-2mu$}\hfill
       \mkern-12mu{\to}\cr
    \noalign{\kern -1pt\nointerlineskip}
      \hfil#1#2\hfil\cr}}}

%

\newcount\figno
\figno=0
\def\fig#1#2#3{
\par\begingroup\parindent=0pt\leftskip=1cm\rightskip=1cm\parindent=0pt
\baselineskip=11pt \global\advance\figno by 1 \midinsert
\epsfxsize=#3 \centerline{\epsfbox{#2}} \vskip 12pt {\bf Fig.\
\the\figno: } #1\par
\endinsert\endgroup\par
}
\def\figlabel#1{\xdef#1{\the\figno}}
\def\encadremath#1{\vbox{\hrule\hbox{\vrule\kern8pt\vbox{\kern8pt
\hbox{$\displaystyle #1$}\kern8pt} \kern8pt\vrule}\hrule}}
\font\cmss=cmss10 \font\cmsss=cmss10 at 7pt

\def\IB{\relax\hbox{$\inbar\kern-.3em{\rm B}$}}
\def\IC{\relax\hbox{$\inbar\kern-.3em{\rm C}$}}
\def\IQ{\relax\hbox{$\inbar\kern-.3em{\rm Q}$}}
\def\ID{\relax\hbox{$\inbar\kern-.3em{\rm D}$}}
\def\IE{\relax\hbox{$\inbar\kern-.3em{\rm E}$}}
\def\IF{\relax\hbox{$\inbar\kern-.3em{\rm F}$}}
\def\IG{\relax\hbox{$\inbar\kern-.3em{\rm G}$}}
\def\IGa{\relax\hbox{${\rm I}\kern-.18em\Gamma$}}
\def\IH{\relax{\rm I\kern-.18em H}}
\def\IK{\relax{\rm I\kern-.18em K}}
\def\IL{\relax{\rm I\kern-.18em L}}
\def\IP{\relax{\rm I\kern-.18em P}}
\def\IR{\relax{\rm I\kern-.18em R}}
\def\Z{\relax\ifmmode\mathchoice
{\hbox{\cmss Z\kern-.4em Z}}{\hbox{\cmss Z\kern-.4em Z}}
{\lower.9pt\hbox{\cmsss Z\kern-.4em Z}} {\lower1.2pt\hbox{\cmsss
Z\kern-.4em Z}}\else{\cmss Z\kern-.4em Z}\fi}

\def\II{\relax{\rm I\kern-.18em I}}

\def\p{\partial}
\def\l{{\lambda}}

\lref\DiFrancescoSS{ P.~Di Francesco and D.~Kutasov,
Phys.\ Lett.\ B {\bf 261}, 385 (1991).
}

\lref\DiFrancescoUD{ P.~Di Francesco and D.~Kutasov,
Nucl.\ Phys.\ B {\bf 375}, 119 (1992) [arXiv:hep-th/9109005].
}

\lref\GinspargIS{ P.~H.~Ginsparg and G.~W.~Moore,
arXiv:hep-th/9304011.
}

\lref\DasHW{
S.~R.~Das, J.~L.~Davis, F.~Larsen and P.~Mukhopadhyay,
``Particle Production in Matrix Cosmology,''
arXiv:hep-th/0403275.
}

\lref\SchomerusVV{ V.~Schomerus, ``Rolling tachyons from Liouville
theory,'' arXiv:hep-th/0306026.
}

\lref\ftt{M.~Headrick, S.~Minwalla and T.~Takayanagi, ``Closed
string tachyon condensation: an overview'', to be published.}
\lref\MooreZV{ G.~W.~Moore, M.~R.~Plesser and S.~Ramgoolam,
``Exact S matrix for 2-D string theory,'' Nucl.\ Phys.\ B {\bf
377}, 143 (1992) [arXiv:hep-th/9111035].
}
\lref\MooreSF{ G.~W.~Moore, ``Double scaled field theory at c = 1,''
Nucl.\ Phys.\ B {\bf 368}, 557 (1992).
}
\lref\MandalTZ{ G.~Mandal, A.~M.~Sengupta and S.~R.~Wadia,
``Classical solutions of two-dimensional string theory,'' Mod.\
Phys.\ Lett.\ A {\bf 6}, 1685 (1991).
}
\lref\SeibergBJ{ N.~Seiberg and S.~H.~Shenker, ``A Note on
background (in)dependence,'' Phys.\ Rev.\ D {\bf 45}, 4581 (1992)
[arXiv:hep-th/9201017].
}

\lref\WittenYR{ E.~Witten, ``On string theory and black holes,''
Phys.\ Rev.\ D {\bf 44}, 314 (1991).
}
\lref\TakayanagiSM{ T.~Takayanagi and N.~Toumbas, ``A matrix model
dual of type 0B string theory in two dimensions,'' JHEP {\bf
0307}, 064 (2003) [arXiv:hep-th/0307083].
}

\lref\DaCunhaFM{ B.~C.~Da Cunha and E.~J.~Martinec,
Phys.\ Rev.\ D {\bf 68}, 063502 (2003) [arXiv:hep-th/0303087].
}

\lref\GrossAY{ D.~J.~Gross and N.~Miljkovic, ``A Nonperturbative
Solution of $D = 1$ String Theory,'' Phys.\ Lett.\ B {\bf 238},
217 (1990);
}
%
\lref\BrezinSS{ E.~Brezin, V.~A.~Kazakov and A.~B.~Zamolodchikov,
``Scaling Violation in a Field Theory of Closed Strings in One
Physical Dimension,'' Nucl.\ Phys.\ B {\bf 338}, 673 (1990);
}
%
\lref\GinspargAS{ P.~Ginsparg and J.~Zinn-Justin, ``2-D Gravity +
1-D Matter,'' Phys.\ Lett.\ B {\bf 240}, 333 (1990).
}
\lref\DasKA{ S.~R.~Das and A.~Jevicki, ``String Field Theory And
Physical Interpretation Of D = 1 Strings,'' Mod.\ Phys.\ Lett.\ A
{\bf 5}, 1639 (1990).
}
\lref\DouglasUP{ M.~R.~Douglas, I.~R.~Klebanov, D.~Kutasov,
J.~Maldacena, E.~Martinec and N.~Seiberg, ``A new hat for the c =
1 matrix model,'' arXiv:hep-th/0307195.
}
\lref\StromingerFN{ A.~Strominger and T.~Takayanagi, ``Correlators
in timelike bulk Liouville theory,'' arXiv:hep-th/0303221.
}

\lref\kkk{ V.~Kazakov, I.~K.~Kostov and D.~Kutasov, `A matrix
model for the two-dimensional black hole,'' Nucl.\ Phys.\ B {\bf
622}, 141 (2002) [arXiv:hep-th/0101011].
}
\lref\GrossUB{ D.~J.~Gross and I.~R.~Klebanov, ``One-Dimensional
String Theory On A Circle,'' Nucl.\ Phys.\ B {\bf 344}, 475
(1990).
}
\lref\sen{ A.~Sen, ``Rolling tachyon,'' JHEP {\bf 0204}, 048
(2002) [arXiv:hep-th/0203211].
}
\lref\mgas{ M.~Gutperle and A.~Strominger, ``Spacelike branes,''
JHEP {\bf 0204}, 018 (2002) [arXiv:hep-th/0202210].
}
\lref\MinicRK{ D.~Minic, J.~Polchinski and Z.~Yang, ``Translation
Invariant Backgrounds In (1+1)-Dimensional String Theory,'' Nucl.\
Phys.\ B {\bf 369}, 324 (1992).
}
\lref\MartinecKA{ E.~J.~Martinec, ``The annular report on
non-critical string theory,'' arXiv:hep-th/0305148.
}
\lref\PolchinskiMB{ J.~Polchinski, ``What is string theory?,''
arXiv:hep-th/9411028.
}

\lref\kms{ I.~R.~Klebanov, J.~Maldacena and N.~Seiberg, ``D-brane
decay in two-dimensional string theory,'' JHEP {\bf 0307}, 045
(2003) [arXiv:hep-th/0305159].
}
\lref\GinspargIS{ P.~Ginsparg and G.~W.~Moore, ``Lectures On 2-D
Gravity And 2-D String Theory,'' arXiv:hep-th/9304011.
}
\lref\hv{ J.~McGreevy and H.~Verlinde, ``Strings from tachyons:
The c = 1 matrix reloaded,'' arXiv:hep-th/0304224.
}
\lref\msy{ A.~Maloney, A.~Strominger and X.~Yin, ``S-brane
thermodynamics,'' arXiv:hep-th/0302146.
}

\lref\bd{ N.~D.~Birrell and P.~C.~W.~Davies, ``Quantum Fields In
Curved Space,'' Cambridge Univ. Pr. (1982).
}

\lref\PolchinskiUQ{ J.~Polchinski, ``Classical Limit Of
(1+1)-Dimensional String Theory,'' Nucl.\ Phys.\ B {\bf 362}, 125
(1991).
} \lref\KlebanovQA{ I.~R.~Klebanov, ``String theory in
two-dimensions,'' arXiv:hep-th/9108019.
}

\lref\NatsuumeSP{ M.~Natsuume and J.~Polchinski, ``Gravitational
Scattering In The C = 1 Matrix Model,'' Nucl.\ Phys.\ B {\bf 424},
137 (1994) [arXiv:hep-th/9402156].
}

\lref\KarczmarekXM{ J.~L.~Karczmarek, H.~Liu, J.~Maldacena and
A.~Strominger, ``UV finite brane decay,'' arXiv:hep-th/0306132.
}

\lref\GaiottoRM{ D.~Gaiotto, N.~Itzhaki and L.~Rastelli, ``Closed
strings as imaginary D-branes,'' arXiv:hep-th/0304192.
}

\lref\PolchinskiJP{ J.~Polchinski, ``On the nonperturbative
consistency of d = 2 string theory,'' Phys.\ Rev.\ Lett.\  {\bf
74}, 638 (1995) [arXiv:hep-th/9409168].
}

\lref\AlexandrovCM{ S.~Alexandrov and V.~Kazakov, ``Correlators in
2D string theory with vortex condensation,'' Nucl.\ Phys.\ B {\bf
610}, 77 (2001) [arXiv:hep-th/0104094].
}

\lref\AlexandrovFH{ S.~Y.~Alexandrov, V.~A.~Kazakov and
I.~K.~Kostov, ``Time-dependent backgrounds of 2D string theory,''
Nucl.\ Phys.\ B {\bf 640}, 119 (2002) [arXiv:hep-th/0205079].
}

\lref\AlexandrovPZ{ S.~Y.~Alexandrov and V.~A.~Kazakov,
``Thermodynamics of 2D string theory,'' JHEP {\bf 0301}, 078
(2003) [arXiv:hep-th/0210251].
}

\lref\AlexandrovQK{ S.~Y.~Alexandrov, V.~A.~Kazakov and
I.~K.~Kostov, ``2D string theory as normal matrix model,'' Nucl.\
Phys.\ B {\bf 667}, 90 (2003) [arXiv:hep-th/0302106].
}
\lref\AlexandrovUH{ S.~Alexandrov, ``Backgrounds of 2D string
theory from matrix model,'' arXiv:hep-th/0303190.
}
\lref\AlexandrovUT{ S.~Alexandrov, ``Matrix quantum mechanics and
two-dimensional string theory in non-trivial backgrounds,''
arXiv:hep-th/0311273.
}
\lref\MinwallaHJ{ S.~Minwalla and T.~Takayanagi, ``Evolution of
D-branes under closed string tachyon condensation,'' JHEP {\bf
0309}, 011 (2003) [arXiv:hep-th/0307248].

}
\lref\DavidVM{ J.~R.~David, M.~Gutperle, M.~Headrick and
S.~Minwalla, ``Closed string tachyon condensation on twisted
circles,'' JHEP {\bf 0202}, 041 (2002) [arXiv:hep-th/0111212].
}

\lref\AdamsSV{ A.~Adams, J.~Polchinski and E.~Silverstein, ``Don't
panic! Closed string tachyons in ALE space-times,'' JHEP {\bf
0110}, 029 (2001) [arXiv:hep-th/0108075].
}
\lref\GutperleMB{ M.~Gutperle and A.~Strominger, ``Fluxbranes in
string theory,'' JHEP {\bf 0106}, 035 (2001)
[arXiv:hep-th/0104136].
}

\lref\SchomerusVV{ V.~Schomerus,
JHEP {\bf 0311}, 043 (2003) [arXiv:hep-th/0306026].
}

\lref\SeibergBJ{ N.~Seiberg and S.~H.~Shenker, ``A Note on
background (in)dependence,'' Phys.\ Rev.\ D {\bf 45}, 4581 (1992)
[arXiv:hep-th/9201017].
}
\lref\ShenkerUF{ S.~H.~Shenker, ``The Strength Of Nonperturbative
Effects In String Theory,'' RU-90-47 {\it Presented at the Cargese
Workshop on Random Surfaces, Quantum Gravity and Strings, Cargese,
France, May 28 - Jun 1, 1990} }
\lref\KarczmarekPV{ J.~L.~Karczmarek and A.~Strominger, ``Matrix
cosmology,'' arXiv:hep-th/0309138.
} \lref\McGreevyEP{ J.~McGreevy, J.~Teschner and H.~Verlinde,
``Classical and quantum D-branes in 2D string theory,'' JHEP {\bf
0401}, 039 (2004) [arXiv:hep-th/0305194].
}
\lref\DabholkarWN{ A.~Dabholkar and C.~Vafa, ``tt* geometry and
closed string tachyon potential,'' JHEP {\bf 0202}, 008 (2002)
[arXiv:hep-th/0111155].
}
\lref\McGreevyKB{ J.~McGreevy and H.~Verlinde, ``Strings from
tachyons: The c = 1 matrix reloaded,'' JHEP {\bf 0312}, 054 (2003)
[arXiv:hep-th/0304224].
}
\lref\cta{ Y.~Okawa and B.~Zwiebach, ``Twisted Tachyon
Condensation in Closed String Field Theory,''
arXiv:hep-th/0403051.
}
\lref\ctb{ M.~Headrick,
arXiv:hep-th/0312213.
} \lref\ctc{ S.~Sarkar and B.~Sathiapalan, ``Closed string
tachyons on C/Z(N),'' arXiv:hep-th/0309029.
}
\lref\bta{ T.~Suyama, ``On decay of bulk tachyons,''
arXiv:hep-th/0308030.
}
\lref\btb{ A.~A.~Tseytlin, ``Magnetic backgrounds and tachyonic
instabilities in closed string  theory,'' arXiv:hep-th/0108140.
} \lref\jlkas{ J.~L.~Karczmarek and A.~Strominger, ``Closed string
tachyon condensation at c = 1,'' arXiv:hep-th/0403169.
}
\def\CIp{${\cal I}^+$}
\def\CIm{${\cal I}^-$}
%
\Title{\vbox{\baselineskip12pt \hbox{hep-th/0405092}
\hbox{SLAC-PUB-10443}}} {\vbox{ \centerline {Hartle-Hawking Vacuum
for}\smallskip
\centerline{ $c=1$ Tachyon Condensation }}} \centerline{Joanna L.
Karczmarek\footnote{$^*$}{Jefferson Physical Laboratory, Harvard
University, Cambridge, MA 02138}, Alexander
Maloney\footnote{$^\dagger$}{SLAC and Department of Physics, Stanford
University, Stanford, CA 94309} and Andrew Strominger$^*$}

\vskip.1in \vskip.1in \centerline{\bf Abstract} Quantum vacua are
constructed for a time-symmetric cosmology describing closed
string tachyon condensation in two-dimensional string theory. Due
to the Euclidean periodicity of the solution, and despite its time
dependence, we are able to construct thermal states at discrete
values of the temperature.  The asymptotic thermal Green functions
and stress energy tensor are computed and found to have an
intriguing  resemblance to those in the Hartle-Hawking vacuum of a
black hole.

\Date{}
\listtoc

\writetoc

\newsec{Introduction}

Soluble string theories in two dimensions provide a useful
laboratory for studying time dependent phenomena such as closed or
open string tachyon condensation and particle production
\refs{\MinicRK\AlexandrovFH \AlexandrovUH \MartinecKA \McGreevyKB
\kms \KarczmarekPV\jlkas\DasHW-\SchomerusVV}. Recently, studies
appeared of particle production during closed string tachyon
condensation \refs{\jlkas,\DasHW}. The ``Alexandrov vacuum'' was
defined in which a tachyon wall distorts the quantum vacuum of the
collective field (describing fluctuations of the fermi surface)
like a reflecting mirror. It was found that when the tachyon
condenses, the mirror accelerates up to \CIp\ and particles are
produced.

In this paper we consider a closely related time-symmetric
cosmology in which the tachyon is condensed in both the far future
and the far past. The Alexandrov vacuum is constructed. Thinking of
the tachyon wall as an analog of the black hole horizon, the
Alexandrov vacuum corresponds to the Boulware vacuum. There is
no energy flux across the tachyon wall (horizon). Observers
comoving with the wall (horizon) will detect no particles. However,
there is a negative Casimir energy for the collective field which
diverges at the wall (horizon).

Further, we define a second, thermal, vacuum state. Thermal
equilibrium is possible at a spectrum of discrete temperatures
despite the time dependence, as in \msy, because of the
imaginary-time periodicity of the solution.  The thermal energy of
the collective field has a positive energy density which diverges
at the tachyon wall due to an infinite blueshift. At a critical
value of the temperature, this divergence is found to exactly cancel 
the negative-energy divergence in the Casimir energy, and the total
stress energy of the collective field becomes finite.
A mathematically identical cancellation was found by
the authors of \DasHW, who include a c-number correction to the
collective field Hamiltonian which is (apparently) not a thermal
energy of the collective field. It would be interesting to better
understand the relation of their observations to ours.

A similar phenomenon occurs for the thermal Hartle-Hawking vacuum
for Schwarzchild black holes. This vacuum can be represented as a
thermal ensemble of Boulware particles built on the Boulware
vacuum. In this description the negative energy Casimir divergence
at the horizon is cancelled by the infinitely blueshifted energy
density of the thermal bath. Although the analogy is not complete,
we accordingly refer to the thermal matrix model vacuum at the
critical temperature as the Hartle-Hawking vacuum.

It seems self evident that these observations are somehow relevant
to the exasperating problem of understanding the role of black
holes in the matrix model (see $e.g.$ \refs{\MandalTZ \WittenYR
\PolchinskiMB\kkk-\AlexandrovPZ}). However, we have nothing
concrete to offer in this direction, and accordingly leave this
problem to the avid reader.

In the next section we adapt the discussion of
\refs{\jlkas,\DasHW}, constructing the Alexandrov vacuum of the
collective field for our time-symmetric cosmology. In section 3 we
construct the thermal state in both the collective field and exact
free fermion pictures. Asymptotic correlators are computed.
Thermality requires that both sides of the fermion barrier are
populated, which in turn entails the 0B interpretation of the
matrix model \refs{\TakayanagiSM, \DouglasUP}. Familiarity with
the free fermion and Das-Jevicki collective field formalisms is
assumed throughout (a useful recent reference is \AlexandrovUT),
and our notation is exactly that of \jlkas.

\newsec{The Alexandrov/Boulware Vacuum}

We are interested in the following solution to the equations of
motion for the fermi surface \eqn\moving{ (x + p - 2\lambda e^t)(x
- p-2\lambda e^{-t}) = 2 \mu~,} where
$x$ and $p$ parameterize the phase space of the free fermions, and
$\mu,\lambda $ are non-negative constants.
This solution describes a
time-reversal invariant cosmological solution in which the tachyon
is condensed on a future portion of \CIp\ and a past portion of
\CIm.
To see this, note that
the fermi surface in this case is a moving hyperbola centered at
\eqn\mmk{(x,p)=(2\lambda\cosh t,2\lambda \sinh t).}
Thus the right branch of the fermi sea is filled at 
time $t=0$, but drains in the
far past and far future, as the hyperbola moves out to large positive $x$.
This differs from the solution considered in \jlkas, which
had tachyon condensation in the future but not in the past.

In this section we will
use the Das-Jevicki formalism to study  collective fluctuations
of this fermi surface, and describe the vacuum states for the
system. In particular, we will construct the ``Alexandrov vacuum''
for \moving, following the discussion of \jlkas.

The hyperbola \moving\ has a second branch centered at the same
point \mmk\ as the first, 
describing the fermi surface spilling over 
from the left hand side of the barrier.
In this section we will consider the bosonic string and ignore the
left side of the fermi sea (which is consistent in perturbation
theory), although it may be easily analyzed. However, the thermal
states considered in the next section necessarily thermally
populate both sides of the barrier, so the left branch of the
fermi sea must be considered.
This will force us to the type 0B interpretation.

\subsec{Fluctuations of the fermi surface}

We wish to describe collective fluctuations of the fermi field in
Das-Jevicki formalism \DasKA, following the conventions and
notation of \jlkas. We may parameterize the right hand
branch of the solution
to \moving\ by introducing a coordinate $\sigma$ that runs from
$-\infty$ to $\infty$, and taking \eqn\stx{\eqalign{ x &=
\sqrt{2\mu} \cosh \sigma + 2\lambda \cosh t \cr p &= \sqrt{2\mu}
\sinh \sigma + 2\lambda \sinh t .}}Introducing the collective
field $\varphi$, which is the difference between the upper and
lower fermi surfaces \eqn\asd{ \varphi = {1\over 2} (p_+ - p_-) }
we find that the background solution \moving\ corresponds to
\eqn\asd{ \varphi_0 = \sqrt{(x-2\lambda \cosh t)^2 - 2\mu} .}
To describe small fluctuations around this background we
define the field $\eta$ by
\eqn\asd{
\varphi = \varphi_0 + \sqrt{\pi} \p_x \eta.}
The dynamics of the small fluctuations are governed by the action \DasKA
\eqn\actionxt{ S=\int dt dx
\Big [ {1\over 2\pi} {(Z_0 + \sqrt{\pi}
\partial_t\eta)^2 \over \varphi_0 + \sqrt{\pi} \partial_x\eta}
-{1\over 6\pi}  (\varphi_0 + \sqrt{\pi}\p_x\eta)^3 + {1 \over \pi}
({1\over 2} x^2 - \mu) (\varphi_0 + \sqrt{\pi} \partial_x \eta)
\Big ]~,}
where \eqn\asd{ Z_0 = \int^x dx' \p_t \phi_0(x') =
-2\lambda \sinh t  ~\varphi_0 .}
The quadratic part of this action
\eqn\quad{\eqalign{
S_2 = {1\over
2} \int {dx dt\over \sqrt{(x-2\lambda \cosh t)^2 - 2\mu} } \big[
(\p_t \eta)^2 &+ 4 \lambda \sinh t (\p_t \eta \p_x \eta) \cr
 &- (x^2 -4x\lambda \cosh t -2\mu + 4 \lambda^2) (\p_x \eta)^2 \big]}}
is that of a scalar field $\eta$ in a curved two dimensional
space. Note from \stx\ that the fermi sea extends from $
\sqrt{2\mu} + 2\l \cosh t < x < \infty $, so it is natural to
impose reflecting boundary conditions on the field $\eta$ along
the ``mirror'' trajectory \eqn\mirror{ x =
\sqrt{2\mu} +2\l \cosh t .}

The action simplifies considerably when written in terms of the
Alexandrov coordinates $(\tau$, $\sigma)$ or $\tau^\pm=\tau\pm
\sigma$, defined by \AlexandrovUT
\eqn\styx{\eqalign{ t=\tau,~~~~~x=
\sqrt{2\mu} \cosh \sigma + 2\lambda \cosh \tau .}}
The full action \actionxt\ becomes
\eqn\actiontau{\eqalign{ \int d\tau d\sigma \Big [ {1\over 2}
((\partial_\tau\eta)^2 - (\partial_\sigma \eta)^2) &- { \sqrt{\pi}
\over 6 \varphi_0^2} (3(\partial_\tau\eta)^2(\partial_\sigma \eta)
+ (\partial_\sigma \eta)^3) \cr &+ \sum_{n=2}^{\infty}
{(-1)^n\over 2}(\partial_\tau \eta)^2
 \Big({\sqrt{\pi} (\partial_\sigma \eta) \over \varphi_0^2}\Big)^n
\Big ], }} with $\varphi_0 = \sqrt{2\mu} \sinh \sigma$.
Note that the quadratic part of the action is precisely that of a scalar field
$\eta$ in flat space $-d\tau^2 + d\sigma^2$.  Moreover, the mirror trajectory
\mirror\ is just the line $\sigma=0$.

We will also consider the ``fermion'' coordinates \eqn\asd{ t^\pm= t
\pm q,~~~~q=\ln x } which are relevant for the spacetime
description of this process near the asymptotic boundaries. The
transformation to the Alexandrov coordinates becomes especially
simple on ${\cal I}^\pm$; on ${\cal I}^+$, as $t^+\to \infty$, \eqn\ttransf{
\tau^+\to\infty, ~~~~~ \tau^- = - \ln
\left(\sqrt{2\over\mu}(e^{-t^-}-\lambda)\right) .}
On ${\cal I}^+$, the mirror is located at the point
$t^- = - \ln \lambda$, where the coordinate transformation
\ttransf\ becomes singular.

\fig{Penrose diagram for spacetime creation/decay via tachyon
condensation.  The dashed line represents the tachyon wall, which
asymptotes to $t^+ = \ln \lambda$ on ${\cal I}^-$ and
$t^- = -\ln \lambda$ on ${\cal I}^+$.}{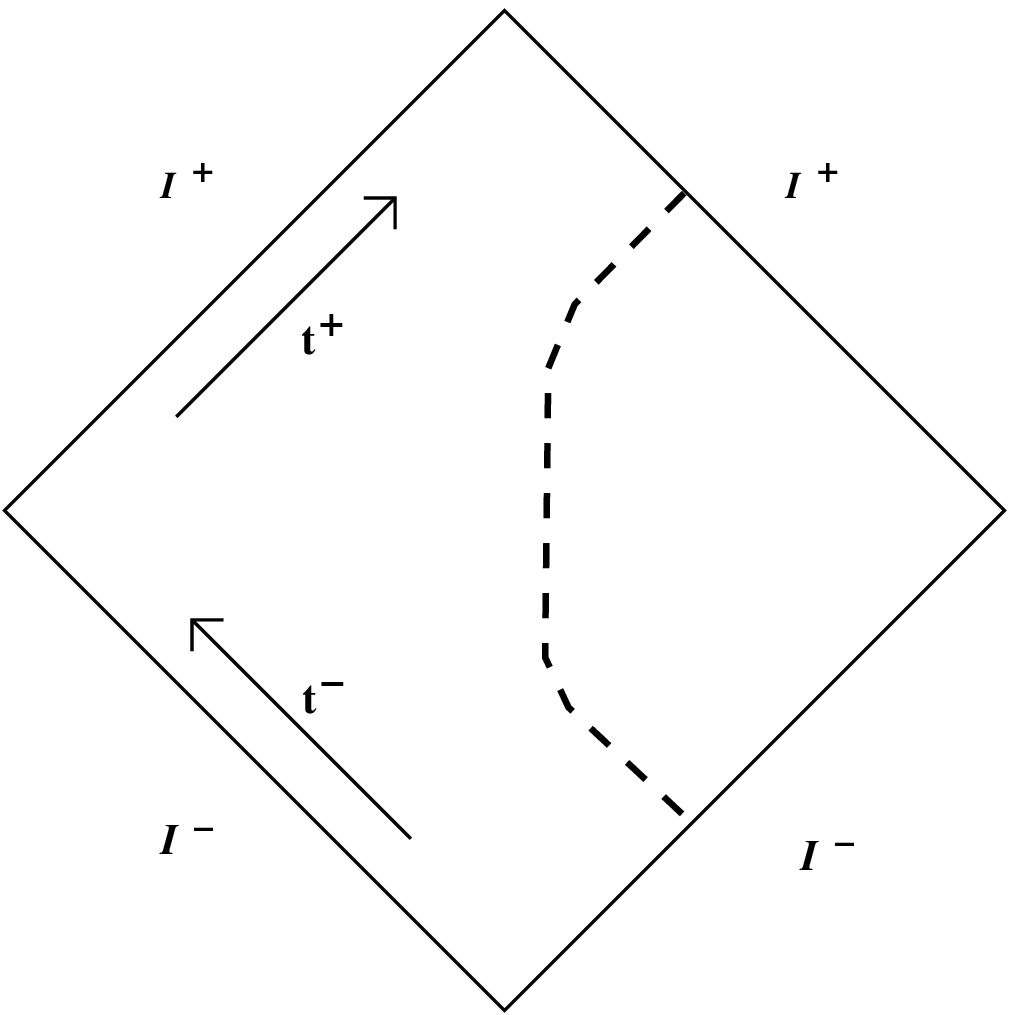}{0.0truein}

\subsec{Vacuum states}

We will now describe vacuum states of the collective field theory
defined in the previous section.  For the rest of this section we
will discuss the free $\eta$ field theory, ignoring the higher
order interactions appearing in \actionxt\ and \actiontau.  The
results of this simplified analysis will turn out to agree with
the full fermi picture, as we will see in the next section.

There are two natural vacuum states $|in\rangle$ and $|out\rangle$,
associated to the positive frequency modes on $\CI^\pm$
\eqn\asd{\eqalign{ u^{in}_\omega &\to {1\over
\sqrt{2\omega} }
              e^{-i\omega t^+},~~~~~{t^-\to-\infty}\cr
u^{out}_\omega &\to {1\over \sqrt{2\omega}}
              e^{-i\omega t^-},~~~~~{t^+\to\infty}
.}}
These states have no particles in the far past and far future, respectively.
They are not equal, but are related by time reversal $t\to -t$.

Following \jlkas, we may also define the Alexandrov
vacuum $|0_A\rangle$ associated to the Alexandrov coordinate plane waves
\eqn\asd{ u^{0}_\omega = {1\over 2\omega} (e^{-i\omega \tau^+} +
e^{-i\omega \tau^-}) .} This state is time reversal invariant, and
differs from $|in\rangle$ and $|out\rangle$.
For example, on ${\cal I}^+$ the outgoing  Alexandrov  modes
depend on $t^-$ as  \eqn\asd{ u^0_\omega \to \left({e^{t^-} -
\lambda\over \sqrt{2\omega}}\right)^{i\omega}. }
The incoming Alexandrov modes take a similar form on ${\cal I}^-$.
We conclude that $|0_A\rangle$ is an excited squeezed state on both
${\cal I}^\pm$.

The energy flux on ${\cal I}^+$ in the Alexandrov state is given by the
Schwarzian of the transformation \ttransf\ between Alexandrov and
$t^-$ coordinates on $\CI^+$,
\eqn\cxg{\eqalign{T_{--}(t^-)&=- {1 \over 12}({\p
\tau^- \over \p t^-})^{3/2} {\p^2 \over \p \tau^{-2}}({\p \tau^-
\over \p t^-})^{1/2}\cr &=  -{\lambda e^{t^-}(2-\lambda e^{t^-})
\over 48(1-\lambda e^{t^-})^{2}}.}}
Note that this diverges at the mirror
$t^- = -\ln \lambda$.
This is the same as the result in \jlkas. However,
unlike \jlkas, a similar nontrivial  Schwarzian gives an energy
flux on ${\cal I}^-$, which diverges at the mirror as well.

\newsec{Thermal States}

In this section we will describe thermal states of the field
theory defined in section 2. In a thermal ensemble both sides of
the barrier will be filled and one must include the second branch of
the hyperbola, 
suppressed in the previous section. While the right branch describes a
draining of the right fermi sea, the left branch describes the fermi
sea spilling over the barrier at late and early times. 
The region $t^-> -\ln \lambda$, left behind by the draining fermi
sea, is filled by the
spilling sea, and all regions of $\cal I$ have a fermi surface
whose fluctuations are the collective field $\eta$.

Once we consider both sides of the fermi sea, we are forced to the
type 0B interpretation \refs{\TakayanagiSM, \DouglasUP} of the
matrix model. In this interpretation, left-right symmetric $\eta$
fluctuations correspond (after a leg pole transformation) to the
NS-NS scalar, while antisymmetric fluctuations correspond to the
RR scalar. In this paper we will present the correlators in
their simplest form, namely in terms of the collective fields for
the left and right fermi surfaces.

Formulae for the 0B leg pole transforms can be found in
\refs{\TakayanagiSM, \DouglasUP}. The leg pole transformation to
spacetime scalars differs for the symmetric and antisymmetric
modes.  These leg pole transformations act as the identity on very
long wavelength fluctuations but smear correlators on the string
scale. In practice they are not analytically computable for the
correlators given here. Additionally, it is not clear how the
standard leg-pole transformations are to be adapted to
non-perturbatively non-trivial situations of the type discussed
herein. One possibility was explored in \jlkas.

\subsec{Stress energy tensor} The tachyon wall trajectory
\eqn\asd{ x = \sqrt{2\mu} + 2\lambda \cosh t }
is invariant under shifts in imaginary time
\eqn\asd{ t\to t+2\pi i n} for any integer $n$. This means \msy\
that we can define a discrete set of thermal states with temperatures
\eqn\temps{ T = {1\over 2\pi n} ,} as we will see in the next subsection.
From \styx\ we can see that  a periodic identification of
imaginary time $t$ leads to a periodic
identification of imaginary $\tau$.  Thus in Alexandrov coordinates these
thermal states are defined by the usual periodicity
\eqn\asd{
\tau = \tau + 2\pi i n.}
This periodic identification in Euclidean space leads to a thermal
energy density \eqn\asd{ T_{--} = {1\over 48 n^2} } in the
$\tau^-$ coordinates.  The Schwarzian to $t^-$ coordinates gives
the energy on $\CI^+$ \eqn\asd{\eqalign{ T(t^-) &= {1\over 48 n^2}
\left( d\tau^-\over dt^- \right)^2 - {1 \over 12}({\p \tau^- \over
\p t^-})^{3/2} {\p^2 \over \p \tau^{-2}}({\p \tau^- \over \p
t^-})^{1/2}\cr &= {1\over 48} + {n^{-2} -1 \over 48(1-\lambda
e^{t^-})^{2}}.}} This is a constant for $n=1$, indicating that
at temperature $T=1/2\pi$
the stress tensor is finite as one approaches the mirror --
in particular, it is constant everywhere on ${\cal I}^+$.
In sections 3.2 and 3.3 we will study this thermal behavior in more
detail by calculating two point functions on $\cal I$.
Finally, in section 3.4 we will discuss the full free fermion
description of these thermal states, and use this formalism to
describe this thermal state in more detail.
%


\subsec{Green functions}

We will now study the Green functions of the pure and thermal states
in the collective field formalism.

First, consider the Alexandrov vacuum $|0_A\rangle$.
In the $\tau$ coordinates
the quadratic part of the action \actiontau\
is just that of a scalar field in flat space with a mirror
located at $\sigma=0$.  The two point function is
\foot{
We consider here correlators of
the derivative of $\eta$, rather than just $\eta$.
This two point function has dimension two, so is the natural object to study
in two dimensions.}
\eqn\asd{\eqalign{
G(\tau,\sigma;\tau',\sigma')
   &= \langle0| \p_{\tau^-}\eta(\tau,\sigma)
\p_{\tau^-{}'}\eta(\tau',\sigma') |0\rangle \cr
   &= {1\over 4\pi} \left(
    {1\over (\tau - \tau' - \sigma + \sigma')^2 }
  + {1\over (\tau - \tau' - \sigma - \sigma')^2 }
\right).}}
The first term is the second derivative of the usual two point function
of a boson in two dimensions.  The second comes from the boundary conditions
at the mirror -- it corresponds to the image charge located at $-\sigma'$.

Thermal Green functions are found by imposing periodicity in
imaginary time. So we may use the standard relation between pure
state and thermal propagators \eqn\summ{\eqalign{
G_{\beta}(\tau,\sigma; \tau',\sigma') &=
  \sum_m G (\tau,\sigma;\tau'+mi\beta,\sigma').}}
This leads to the thermal propagator
\eqn\asd{\eqalign{
G_{\beta}(\tau,\sigma; \tau',\sigma')
&= {\pi \over 4 \beta^2} \left(
{1\over \sinh^2 \pi (\tau-\tau'-\sigma+\sigma')/\beta }+
{1\over \sinh^2 \pi (\tau-\tau'-\sigma-\sigma')/\beta }
\right).
}}

We can now transform this back to the $(t,x)$ coordinates.
For $\beta=2\pi n $ we can perform the sum \summ\ in either
the $\tau^\pm$ or $t^\pm$ coordinates, since the coordinate
$\sigma$ defined by \styx\
is invariant under the shift $t\to t+i\beta$.  The thermal
propagator becomes
\eqn\grf{\eqalign{
G_\beta (t,x;t',x') &= \langle 0| \p_{t^-}\eta \p_{t^-{}'}\eta |0\rangle \cr
&= J J'
{\pi \over 4 \beta^2} \left(
{1\over \sinh^2 \pi (t-t'-\sigma+\sigma')/\beta }+
{1\over \sinh^2 \pi (t-t'-\sigma-\sigma')/\beta }\right) \cr}}
where
\eqn\asd{\eqalign{
\sigma &= \cosh^{-1} {x-2\l \cosh t\over \sqrt{2\mu}} .}}
We are considering here the two point function of
$\p_{t^-}\eta $ rather than $\p_{\tau^-}\eta$, so we must include
the Jacobian $J$ evaluated at constant $t^+$
\eqn\asd{
J = \left( {\p \tau^-\over \p t^-}
\right)^{-1}_{t^+}
.}

This Green function takes a particular simple form on $\scri^+ $.
First note that $\sigma$ is
only well defined in the region
of $\scri^+$ with
\eqn\region{-\infty < t_- < -\ln \l,}
where
\eqn\asd{
\sigma\to \ln\sqrt{2/\mu} + t + \ln (e^{-t^-} - \l) \to \infty.}
The second term in \grf\ then vanishes as $e^{-t^+}$ , and the first term
gives
\eqn\therm{\eqalign{
G_\beta &= J J'
{\pi \over 4 \beta^2 \sinh^2 \pi\ln \left(
e^{-t^-{}'} - \l\over e^{-t^-} - \l\right)/\beta }
.}}
When $n=1$, so that $\beta=2\pi$, several terms cancel and this becomes the
usual thermal propagator in two dimensions
\eqn\ttherm{
G_{\beta=2\pi} = {1\over 16 \pi \sinh^2 (t^- - t^-{}')/2}
.}
We conclude that the physical tachyon correlation functions on the
boundary with both points in the region \region\ are precisely
thermal.
The case where both
points are not in this region is more interesting, as we will now
discuss.

\subsec{Correlators on all of $\CI^+$}

To study correlations on the rest of ${\cal I}^+$,
it is useful to
define a coordinate $r$ which covers all of $\CI^+$ -- we will take
\eqn\asd{\eqalign{
r &= -\exp(-\tau^+)~~~{\rm as}~~ \tau^- \rightarrow +\infty,\cr
r &= \exp(-\tau^-)~~~{\rm as}~~ \tau^+ \rightarrow+\infty
.}}
We can define a similar coordinate $R$ which is related to
the fermion coordinates $t^{\pm}$,
\eqn\Rcoord{\eqalign{
R &= -\exp(-t^+) ~~~{\rm{as}}~~ t^- \rightarrow +\infty~,\cr
R &= \exp(-t^-) ~~~{\rm{as}}~~ t^+ \rightarrow +\infty~.
}}
So $R\sim x e^{-t}$ as $t \rightarrow +\infty$.
These coordinates are useful because the coordinate change from Alexandrov
to fermion coordinates is particularly simple: on ${\cal I}^+$,
 $R= r + \lambda$.
For $n=1, \beta = 2\pi$,
the thermal two point function on the left hand component of
${\cal I}^+$ (where $\tau^-\to\infty$) reduces to
\eqn\GLL {G_{\beta=2\pi}(\tau^+, \tau'^+) \equiv
\langle 0| \p_{\tau^+}\eta(\tau^+)
      \p_{\tau^+{}'}\eta(\tau'^+) |0 \rangle =
{1 \over 16\pi} {1 \over \sinh^2\big((\tau^+ - \tau'^+)/2\big)}
.}
On the right hand part of ${\cal I}^+$ ($\tau^+\to\infty$),
\eqn\GRR {G_{\beta=2\pi}(\tau^-, \tau'^-) \equiv
\langle 0| \p_{\tau^-}\eta(\tau^-)
      \p_{\tau^-{}'}\eta(\tau'^-) |0 \rangle =
{1 \over 16\pi} {1 \over \sinh^2\big((\tau^- - \tau'^-)/2\big)}
.}
The correlator when one point lies on the right hand component of
$\CI^+$ and the other lies on the left hand component vanishes,
\eqn\GRL {G_{\beta=2\pi}(\tau^+, \tau'^-) \equiv
\langle 0| \p_{\tau^+}\eta(\tau^+)
      \p_{\tau^-{}'}\eta(\tau'^-) |0 \rangle = 0.}
These three formulas are compactly written in the $r$ coordinate as
\eqn\Grr {G_{\beta=2\pi} (r, r') = {1 \over 16\pi} {(r-r')^2 \over r r' }
~\Theta(r r').}
Transforming to the fermion $R$ coordinate,
using the Jacobians $\p r / \p \tau^- = \p r / \p \tau^- = -r$
and $\p R / \p t^- = \p R / \p t^- = -R$, the correlator becomes
\eqn\GRR {G_{\beta=2\pi}(R, R') = {1 \over 16\pi} {(R-R')^2 \over R R'}
~\Theta((R-\lambda)( R'-\lambda)).}
This reproduces the thermal correlator \ttherm\ when both points
lie on the same component of $\CI^+$,
and allows one to compute correlators between points in
different components of $\CI^+$.
For example, in the fermion coordinates the correlator
between one point on the right hand side of ${\cal I}^+$
and another on the left hand side of
${\cal I}^+$ (with $t^-{}'\to \infty$) does not vanish.  In particular,
if the point on the right hand side of $\CI^+$ lies in the region
with $t^- > -\ln(\lambda)$
then
\eqn\Gxx {G_{\beta=2\pi}(t^- > -\ln(\lambda), t'^+ ) = - {1 \over 16\pi}
{1 \over \cosh^2\big((t^- - t'^+)/2\big)}.}

These correlators are exactly thermal, but with a reflecting
mirror inserted at $t^-=-\ln (\lambda)$ on the right hand side of
${\cal I}^+$. Correlators for the spacetime axion and tachyon of
the 0B theory may be obtained from these via a leg pole
transformation of the left-right antisymmetric and symmetric
linear combinations of $\eta$, but we shall not give the explicit
expressions.

\subsec{Exact free fermion picture}

In this subsection, we
will show, using the  free fermion description,
that the thermal correlators derived above in the
collective field formalism are actually exact.  We will also
give a precise meaning to the preceding statement that there
exist thermal states with temperature $T=1/2\pi n$.

The difference between the quantum state describing the original
static fermi sea and the state describing the filling and draining
of the fermi sea is too large to be described as a state in the
Hilbert space of the original theory \SeibergBJ. In the language
of \SeibergBJ, it involves non-normalizable modes. Hence it is
associated to a new Hamiltonian rather than to a semiclassical
quantum state in the theory governed by the old Hamiltonian.
This new, time-dependent Hamiltonian can be written as
\eqn\wws{H_\lambda=\half(p^2-x^2) +\lambda e^t (x-p) +\lambda
e^{-t} (x+p) ~.}

Consider the new phase space
coordinates \eqn\ydef{
(y, p_y) \equiv (x - 2\lambda \cosh t, p - 2\lambda
\sinh t),} which obey the usual equations of motion $\dot
y = p_y,~ \dot p_y = y$.  In these coordinates the
Hamiltonian is $(p_y^2 - y^2)/2$ and the fermi
surface \moving\ is given by $(y^2 - p_y^2) = 2 \mu$. We know
the quantum theory corresponding to this classical limit -- it is
simply the theory of free fermions with potential $-y^2/2$, with
all states up to $\mu$ below the top filled.
Thus exact
correlators at finite $\lambda$ are related to the correlators in the
usual $\lambda=0$ theory (which can be computed as in \MooreSF)
by the coordinate shift \ydef.
This procedure reproduces the
results derived above in the collective field approximation. The
argument is a straightforward adaptation of the one given in
\jlkas\ and will not be repeated here.

We may also use the free fermion picture to understand the physical
meaning of these thermal states.  Typically, thermal
states are only defined for time independent systems, where
correlators calculated in a compact Euclidean space
$t\sim t+ i \beta$ agree with those computed from a density matrix
$\rho=e^{-\beta H}$.  In our case, the Hamiltonian \wws\
depends on time, so it is not clear which density matrix
is associated to the Euclidean identification $t\sim t+2\pi i n$.
In fact, as we will now show, the density matrix is
\eqn\asd{\rho = e^{- 2 \pi n H_0}}
where $H_0$ is the time independent part of the Hamiltonian,
and will be defined shortly.

To see this, we will use
the second quantized language, where
the fermion field $\Psi(x,t)$ can be expanded as
\eqn\Psimodes{
\Psi(x,t) = \sum_{\omega, s} a_\omega^{(s)} \psi_\omega^{(s)}
(x-2\lambda\cosh t,t) .}
The wave functions $\psi_\omega^{(s)}(y,t)$, for $s=\pm1$, are the
symmetric/antisymmetric solutions to the one particle Schrodinger
equation at energy $\omega$
\eqn\schr
{ -{1\over 2}(\partial_y^2 + y^2) \psi_\omega^{(s)}(y,t)
= -i \partial_t  \psi_\omega^{(s)}(y,t) = \omega \psi_\omega^{(s)}(y,t)
.}
The second quantized Hamiltonian
\eqn\H{
H(t) = \int dx \Psi^\dagger(x,t)\Big [
{1 \over 2} (\partial_x^2 + x^2)
+\lambda e^t (x+i\partial_x) +\lambda e^{-t} (x-i\partial_x)
-2\lambda^2 \cosh^2 t
\Big ]\Psi(x,t)
}
gives the Schrodinger evolution of $\Psi$
\eqn\Schr
{-i \partial_t \Psi(x,t) = [H(t), \Psi(x,t)]
.}
We define now the time independent portion of the
Hamiltonian, given by (up to a zero-point energy)
\eqn\Hzero{ H_0 = \sum_{\omega, s}
\big(a_\omega^{(s)}\big)^\dagger a_\omega^{(s)} \omega .}
The operator $H_0$ generates the following shift in imaginary time
\eqn\shift { e^{2 \pi n H_0} \Psi(x,t) e^{-2 \pi n H_0} = \Psi(x,
t+2\pi i n) } since each wave function $\psi_\omega$ obeys
\eqn\zzz{ {e^{-2 \pi n \omega}}
\psi_\omega(x-2\lambda\cosh t,t)  = \psi_\omega(x-2\lambda\cosh
t',t')\Big|_{t' = t + 2\pi i n} .} Therefore, the
Heisenberg-picture density matrix is given in equation
\asd. Correlation functions in this density matrix agree
with those computed by analytic continuation from a periodically
identified Euclidean theory, as in \msy.

\centerline{\bf Acknowledgements}
This work was supported in part by DOE grants DE-FG02-91ER40654 and
DE-AC03-76SF0015.
We are grateful to J. Maldacena and S. Shenker for
useful conversations.

\listrefs
\end

The Hamiltonian of the system is \eqn\asd{ H_\lambda = {1\over 2}
(p^2 - x^2) - \l e^t (x-p) - \l e^{-t} (x+p) .} In terms of the
shifted coordinates \eqn\asd{ y=x-\l\cosh t,~~~~p_y=p-\l\sinh t }
the Hamiltonian is trivial and independent of $\lambda$ \eqn\asd{
H =  {1\over2} (p_y^2 - y^2) .}

Finally, let us ask about the relationship between the $|in>$ and
the $|out>$ vacua.  Noting that \eqn\asd{\eqalign{ e^{t^+} &\to
\lambda + \sqrt{\mu\over2} e^{\tau^+},~~~~~t^-\to-\infty\cr
e^{-t^-} &\to \lambda + \sqrt{\mu\over2}
e^{-\tau^-},~~~~~t^+\to\infty }} and using the fact that a general
solution to the wave equation takes the form $f(\tau^+)+f(\tau^-)$
in the bulk, if follows that the full mode solutions are
\eqn\asd{\eqalign{ u^{in}_\omega &= \left(\lambda+\sqrt{\mu\over2}
e^{\tau^+}\right)^{-i\omega} + \left(\lambda+\sqrt{\mu\over2}
e^{\tau^-}\right)^{-i\omega} \cr u^{out}_\omega &=
\left(\lambda+\sqrt{\mu\over2} e^{-\tau^+}\right)^{i\omega} +
\left(\lambda+\sqrt{\mu\over2} e^{-\tau^-}\right)^{i\omega} .}}
These are very similar, but appear to differ, indicating the the
$|in>$ and $|out>$ vacua are not the same (CHECK?  it would be
interesting if they were the same).  Note that time reversal acts
as $\tau^\pm\to -\tau^\mp$ along with complex conjugation,
indicating, as we expect, that $|in>$ and $|out>$ are related by
time reversal.

In the region $-\ln \l < t^- < \infty$ the argument of $\cosh^{-1}$ in the
definition of $\sigma$ is negative.  Expression
\grf\ still defines, by analytic continuation the correlator as a function
of $x,t$, as long as we are careful  with the branch cuts.
We may still formally write
\eqn\asd{
\cosh^{-1} z = \ln (\sqrt{z^2 -1} + z)  = \pi i + \cosh^{-1} (-z)}
so that
\eqn\asd{
\sigma \to \pi i + \ln \sqrt{2/\mu} + t + \ln (e^{-t^-} -\l)
}
This leads to thermal correlators \therm\ when both points are in the
upper region.  When the two points are in different regions of
$\scri^+$ the computation is trickier.

the two regions.  CHECK
\newsec{Ignore -- Danger: Whitaker functions}
By the above argument, we need only study thermal states of the
free $\lambda=0$ theory. We wish to calculate the Green function
\eqn\asd{ G_\beta(y,t;y',t') = <\psi^\dagger (y,t) \psi
(y',t')>_{\beta = 2\pi n} } at finite temperature $T = 1/2\pi n$.
$G$ is defined to be periodic in imaginary time and solves the
wave equation \eqn\asd{ (i\p_t + \p_y^2 - {y^2 /2 } - \mu) G =0 }
along with a similar relation for $y'$.  Expanding into modes, we
have \eqn\asd{ G = \sum_\nu {1\over 2\nu} e^{-i\nu (t-t')/n}
G_{\nu/n+\mu} (y',y) } where $G_a$ is the spatial Green function
satisfying \eqn\asd{ (\p_y^2 - y^2 /2 - a) G_a =0 } and a similar
equation for $y'$. Letting $\psi^\pm_a (y)$ be the even (odd)
solution to the wave equation \eqn\asd{ (\p_y^2 - y^2/2 - a)
\psi=0 } we find that the delta function normalized solution is
\eqn\asd{ \psi^\pm_a = {1\over \sqrt{4\pi(1+e^{2\pi a})^{1/2}}}
(W(a,y) \pm W(a.-y)) } where $W$ is the parabolic cylinder
function.  Then \eqn\asd{ G_a = (\psi^+_a (y') )^* \psi^+_a (y) +
      (\psi^-_a (y') )^* \psi^-_a (y)
.}

We wish to study these Green functions when both points are on
$\scri^+$.  This means taking $t=t'\to \infty$, so that $t^+{}' =
t^+ + t^--t^-{}'$, $x=e^{(t^+ - t^-)/2} $ and \eqn\asd{\eqalign{
y&= e^{(t^++t^-)/2} (e^{-t^-}-\l)\cr y'= e^{(t^+ + t^-)/2}
(e^{-t^-{}'} - \lambda) }} become either large and positive or
large and negative. First, note that when $y\to \infty$ the
cylinder functions become \eqn\asd{\eqalign{ \psi^\pm_a(y) &\to
{1\over \sqrt{2\pi y(1+e^{2\pi a})^{1/2}}} (k^{1/2} \cos\theta \pm
k^{-1/2} \sin \theta) \cr k &=\sqrt{1+e^{2\pi a}} - e^{\pi
a},~~~~~ \theta = y^2/4 - a \ln y + \pi/4 + \arg\Gamma(1/2 +
ia)/2 .}} The behavior as $y\to -\infty$ is the same, except that
$\psi^-$ has an minus sign in front of it.  Let's consider the two
point function where both $y$ and $y'$ are positive.  Then and
\eqn\asd{\eqalign{ G_a &= {1\over 2\pi \sqrt{1+e^{2\pi a }}}
{1\over \sqrt{y y'}} \left[(k+k^{-1}) \cos (\theta - \theta') / 2
+(k-k^-1) \cos(\theta + \theta')/2 \right] \cr &= {1\over \sqrt{y
y'}} \left[ \cos (\theta - \theta')/2 +{\cos(\theta + \theta')/2
\over 2\pi \sqrt{1+e^{-2\pi a}}} \right].}} When $y'$ is negative,
we get \eqn\asd{\eqalign{ G_a &= {1\over 2\pi \sqrt{1+e^{-2\pi a}
} \sqrt{y y'}} \left[\sin (\theta - \theta') / 2 + \sin(\theta +
\theta')/2 \right] .}}

Let's consider the pure state.  For $y'$ positive the first term
contributes to the Green function \eqn\asd{\eqalign{ G &= {1\over
\sqrt{yy'}} \int_0^\infty d\nu
  \cos ( (y^2 - y'^2 )/4 - (\nu+\mu)  \ln (y/y'))\cr
&= Re {e^{i [(y^2-y'^2)/4 - \mu \ln (y/y')]} \over \sqrt{yy'}
\ln (y/y')} .}} Here I've added a small imaginary part to $y$ to
get the integral to converge.  Up to a phase, this is \eqn\asd{ .}
For $y'$ negative the first term is \eqn\asd{\eqalign{ G&= Im {i
e^{i [(y^2-y'^2)/4 - \mu \ln (y/y')]} \over 2\pi \sqrt{yy'}}
\int_0^\infty d\nu
             {e^{i \nu \ln (y/y') } \over \sqrt{1+e^{-2\pi (\nu + \mu)}}}
}} This is dominated by the small $\nu$ piece, which gives an
answer similar to the previous one.

In the thermal state, we have for $y'$ positive \eqn\asd{\eqalign{
G &= {1\over \sqrt{yy'}}  \sum_\nu
  \cos ( (y^2 - y'^2 )/4 - (\nu/n+\mu)  \ln (y/y'))\cr
&= Re {e^{i [(y^2-y'^2)/4 - \mu \ln (y/y')]} \over \sqrt{yy'}
(1-(y/y')^{1/n}) } }} and for $y'$ negative \eqn\asd{\eqalign{ G&=
Im {i e^{i [(y^2-y'^2)/4 - \mu \ln (y/y')]} \over 2\pi
\sqrt{yy'}} \sum_\nu
             {e^{i \nu/n \ln (y/y') } \over \sqrt{1+e^{-2\pi (\nu/n + \mu)}}}
}} which turns out to be exactly the same in the small $\nu$
region, at least for $n=1$.


*****************8 Let's see how.

Let's study the transformation between $(x,t)$ and $(y,t)$
coordinates more carefully.  First, recall that the we
parameterize points on the right (left) components of $\scri$ by
combinations like $t_R^\pm = t \pm \ln(y)$ ($ t_L^\pm = t \pm \ln
(-y)$). Note that as $t^+_R\to\infty$ we have \eqn\asd{ t+\ln y
\to (t^+ + t^-) + \ln (e^{-t^-}-\l)\to \infty,~~~~~ t-\ln y \to -
\ln (e^{-t^-}-\l)}. So the region on $\scri^+_R(x)$ with $-\infty<
t^-<-\ln \l$ maps to all of $\scri^+_R(y)$.  And when $-\ln \l <
t^- < \infty$ we have \eqn\asd{ t+\ln (-y) \to (t^+ + t^-) + \ln
(\l-e^{-t^-})\to \infty,~~~~~ t-\ln (-y) \to - \ln (\l -
e^{-t^-})} so that this region of $\scri^+_R(x)$ maps to the part
of $\scri^+_L(y)$ with $-\ln\l < t-\ln(-y)<\infty$. What about
$\scri^+_L(x)$? This requires taking $t^+_L = t+\ln (-x)\to
\infty$, so that \eqn\asd{ t+\ln(-y) \to (t^+_L + t_L^-)+ \ln (\l
+ e^{- t^-_L}) \to \infty,~~~~~ t-\ln(-y) \to - \ln (\l +
e^{-t^-_L}) .} We conclude that all of $\scri^+_L(x)$ maps to the
part of $\scri^+_L(y)$ with $-\infty <t-\ln(-y)<-\ln \l $. Similar
statements can be made about $\scri^-$.

>From this point of view the results of the previous section
follow...